\documentclass[twocolumn,prl,aps,showpacs]{revtex4}
\usepackage{graphicx}
\usepackage{epstopdf}
\begin{document}

\title{Observation of Reduced Three-Body Recombination in a Fermionized 1D Bose Gas}

\author{B.~Laburthe~Tolra}
\altaffiliation[Permanent address: ] {\protect{Laboratoire de Physique des Lasers (UMR 7538),  Universit\'{e} Paris 13.}}
\author{K.~M.~O'Hara}
\author{J.~H.~Huckans}
\author{W.~D.~Phillips}
\author{S.~L.~Rolston}
\altaffiliation[Permanent address: ] {Department of Physics, University of Maryland, College Park, MD.}
\author{J.~V.~Porto}
\altaffiliation[To whom correspondence should be addressed: ]{trey@nist.gov}

\affiliation{National Institute of Standards and Technology,
Gaithersburg, MD 20899}

\date{\today}

\begin{abstract}
We investigate correlation properties of a one-dimensional
interacting Bose gas by loading a magnetically trapped
$^{87}$Rb Bose-Einstein condensate into a deep two-dimensional
optical lattice. We measure the three-body recombination rate for
both the BEC in the magnetic trap and the BEC loaded into the
optical lattice. The recombination rate coefficient is a factor of seven smaller in the lattice, which we interpret as a
reduction in the local three-body correlation function in the 1D case. This is a
signature of correlation intermediate between that of
the uncorrelated phase coherent 1D mean-field regime and the strongly
correlated Tonks-Girardeau regime.
\end{abstract}

\maketitle

The majority of experiments with quantum degenerate gases have been performed in the weakly interacting limit, on Bose-Einstein
condensates (BECs) characterized by long-range phase coherence and
well described by the mean-field Gross-Pitaevskii (GP)
equation~\cite{stringari}.  While the success of the GP equation
in accounting for many experimental results has been
spectacular, it has also led to the search for physics beyond mean-field theory. As in condensed matter physics, there is now great
interest in highly correlated systems, where mean-field
approaches are inapplicable. Progress toward such correlated
systems includes the recent observation of the Mott-insulator
transition in BECs loaded into optical lattices~\cite{Greiner02}
and the use of Feshbach resonances to increase interactions
between atoms~\cite{bosanova}.  Here we present evidence of strong
correlations in a 1D degenerate Bose gas as reflected in a
reduction of three-body recombination.

The role of fluctuations and correlations in Bose gases increases
with reduced dimension. In homogenous systems, BEC is only
possible in 3D. In 2D, a Kosterlitz-Thouless transition occurs,
and in 1D there is no finite temperature
transition~\cite{phasetransition}. By contrast, BEC is possible in
1, 2 and 3D for trapped systems~\cite{Bagnato}.  Trapped 1D
systems with $\delta$-function repulsive interactions are
particularly interesting, in that for high density the ground
state is a condensate, while in the low density limit the ground
state is a highly correlated state known as a Tonks gas~\cite{LL}.
This ground state is an example of ``fermionization,'' where the
repulsive interactions mimic the Pauli exclusion principle.
Indeed, the low energy excitation spectrum and the correlation
functions are identical to those of non-interacting fermions, and
the many-body wave function of the Bose gas is equal to the
absolute value of the fermionic wave function~\cite{girardeau}.

For such a 1D Bose gas, the degree of correlation depends on the
ratio between two energies: the repulsive energy of uncorrelated
atoms at a given density, $E_{\mbox{\scriptsize unc}} = g
n_{\mbox{\scriptsize 1D}}$, and the quantum kinetic energy needed
to correlate particles by localizing them with respect to each
other on the order of the mean inter-particle distance $d$,
$E_{\mbox{\scriptsize cor}} = \hbar^{2}/2 m d^{2}$. Here, $g$ is
the strength of the $\delta$-function interaction, $m$ is the atomic mass, and $n_{\mbox{\scriptsize 1D}} = 1/d$ is the 1D density. A
single parameter $\gamma \equiv E_{\mbox{\scriptsize unc}}/4
E_{\mbox{\scriptsize cor}} =mg/ 2 n_{1 \mbox{\scriptsize D}}
\hbar^{2}$ entirely characterizes a homogeneous 1D gas with
repulsive short range interactions. Ref.~\cite{LL} provides the
exact eigenstate  solutions for all $\gamma$. Recently, these results have been extendend to harmonically trapped gases, addressing, {\it e.g.}, the excitation spectrum~\cite{menotti02}, the shape of the trapped gas~\cite{dunjko01}, and correlation properties~\cite{gangardt03}.

The many-body ground state has two limiting forms. In the
Tonks-Girardeau (TG) regime, where $\gamma \gg 1$, the ground
state becomes correlated in order to minimize the interaction
energy and the bosons become impenetrable, behaving like fermions
as described in \cite{girardeau}. The second and higher-order local correlation functions $g_{i}$ vanish~\cite{kheruntsyan}, meaning that no more
than one particle can be found at a given
position.  On the other hand, in the mean-field
(MF) regime when $\gamma \ll 1$, the GP equation
describes the system well. In this regime the healing length, $l_h
= \hbar/\sqrt{m g n_{\mbox{\scriptsize 1D}}}$, is much larger than
the mean inter-particle distance.  Note the counter-intuitive result
that the system reaches the correlated regime for {\em low} 1D
densities, contrary to the 3D case where $n_{3 \mbox{\scriptsize
D}}a_s^3 = (\gamma/2 \pi )^3 \gg 1$ corresponds to the correlated regime. (Here $a_s$ is the zero-energy 3D scattering length~\cite{vankempen02}, and $\gamma$ is the appropriate energy ratio in 3D.)

To probe correlations we measure three-body recombination rates
(proportional to the local third-order correlation function $g_{3}$) of
1D gases produced in a 2D optical lattice. This technique was used in Ref.~\cite{burt97}  to demonstrate that there is a reduction
of $g_{3}$ in a 3D BEC by a factor of six compared to a thermal
gas. We observe a further reduction of three-body
recombination in a 1D gas compared to the 3D BEC situation. Even
though $\gamma \simeq 0.5$ for our system, which is far from the
TG regime, this is a signature that the correlations are
significant due to the ``fermionization'' of the particles.

We realize a 1D gas by confining a 3D gas sufficiently tightly in two
directions that the radial confinement energy $\hbar
\omega_{\bot}$ is much larger than all other relevant energies in
the system: $E_{\mbox{\scriptsize unc}}$, $E_{\mbox{\scriptsize
cor}}$, $k_{B} T$, and the axial trapping energy $\hbar
\omega_{z}$.  Since $a_s$ is much smaller than $a_\bot = \sqrt{\hbar/m \omega_\bot}$ ($a_s/a_\bot \simeq 0.1$ in our system), the atom-atom interaction strength is largely determined by $a_s$, with only a small correction due to confinement~\cite{olshanii98,eite}: $a_{\mbox{\scriptsize eff}} = a_s/(1-1.46\ a_s/\sqrt{2} a_{\bot})$. There is no excitation in the radial direction and by integrating over
the radial coordinates, one can show~\cite{olshanii98} that the
system is formally equivalent to a true 1D gas with interaction
strength $g = 4 \hbar^{2} a_{\mbox{\scriptsize eff}} / m a_{\perp}^{2}$, so that $\gamma = 2 a_{\mbox{\scriptsize eff}} /(n_{\mbox{\scriptsize 1D}} a_{\bot}^2)$. The 3D density
is related to the effective 1D density by $n_{1 \mbox{\scriptsize
D}} = 1/d = \pi a_{\bot}^{2} n_{3 \mbox{\scriptsize D}}$.

Our approach is to load a BEC into the ground state of a deep 2D
optical lattice so that the BEC is divided into an array of
independent 1D quantum gases, each tubular lattice site acting as a highly
anisotropic trap. Our experimental apparatus has been described
elsewhere \cite{peil03}. We achieve BEC with up to
$N_0 = 5 \times 10^{5}$ atoms in
the $(F,m_F)=(1,-1)$ hyperfine state of $^{87}$Rb (for which $a_s = 5.313$~nm~\cite{vankempen02}). A Ioffe-Pritchard trap
confines the atoms with initial ``tight'' trap frequencies of
$\nu_{x}=\nu_{z}=210$~Hz, and $\nu_{y} =8.2$~Hz, giving a peak
atomic density of up to $3 \times 10^{14}$ cm$^{-3}$. Before
applying the optical lattice, we adiabatically lower $\nu_{x}$ and
$\nu_{z}$ to a ``weak'' trap frequency of 28~Hz, resulting in peak
densities of $\sim 5 \times 10^{13}$ cm$^{-3}$.

We create a 2D optical lattice from two independent,
retro-reflected 1D lattices which lie in the $xy$-plane and
intersect at an angle of 80 degrees.  The independent 1D lattices
are detuned from each another by 5~MHz. All beams derive from a
Ti:sapphire laser operating at $\lambda=810.08$~nm (detuned below
both $5S \rightarrow 5P$ transitions at 795~nm and 780~nm), and
the polarizations of the lattice beams are in the $xy$-plane. Each
1D lattice is measured~\cite{depth} to be 29(1)~$E_{R}$ deep
(where $E_{R}=h^2/2 m \lambda^{2}$)~\cite{stddev}.  At each
lattice site the ground state of the radial motion is well
approximated by a gaussian wavefunction with $a_{\bot} = 58.5(5)
{\mathrm {nm}}$ corresponding to an effective $\omega_{\bot} / 2
\pi =$~33.8(6)~kHz. By observing dipole oscillations following a
sudden, brief displacement of the trap center, we measure the
axial frequency along the tubes to be $\omega_z /2 \pi =$ 55.9(6)
Hz. This frequency results from the combined effect of the
magnetic trap and the dipole potential of the lattice beams along
the tubes.  To load the atoms into the lattice, the laser light is
increased over 200~ms with an approximately half gaussian shape
(rms width 70 ms), which is adiabatic with respect to all
vibrational excitations. We estimate that the interaction-free
tunnelling time from one lattice site to the next for a 29~$E_{R}$
lattice is $\simeq$150 ms. Although this is shorter than the time
of the experiment (up to 12~s), it corresponds to an energy much
smaller than the interaction energy in the tubes, and should not
modify the local 1D correlation properties~\cite{pedri03}.

To measure the reduction of $g_3$ due to correlations, we observe
the corresponding reduction in the three-body recombination rate
coefficient. The local three-body recombination rate (in either 1D
or 3D) is proportional to the cube of the local density. For
$^{87}$Rb, it is known that two-body
losses~\cite{burt97,soding99}, including photoassociation at 810~nm~\cite{carl}, are very small~\cite{K2note}. Our model, therefore, includes only one-body and three-body processes so that the total number of atoms $N$ decays according to
\begin{equation}\label{lossrate}
\frac{dN}{dt} = -K_{1}\  N -\ \int  K_{3}^{{ \mbox{\scriptsize
1D}}}\ n_{ \mbox{\scriptsize 3D}}^3\ dV.
\end{equation}
We account for atomic redistribution during decay through
the evolution of the density profile. Determination of the
three-body recombination rate coefficient
$K_{3}^{{\mbox{\scriptsize 1D}}}$ requires an accurate estimate of
the density, which we ascertain from a measurement of the number
of trapped atoms as a function of time, along with a determination
of the size and shape of the atom cloud.

\begin{figure}[t]
\begin{center}
\includegraphics[width=3.24in]{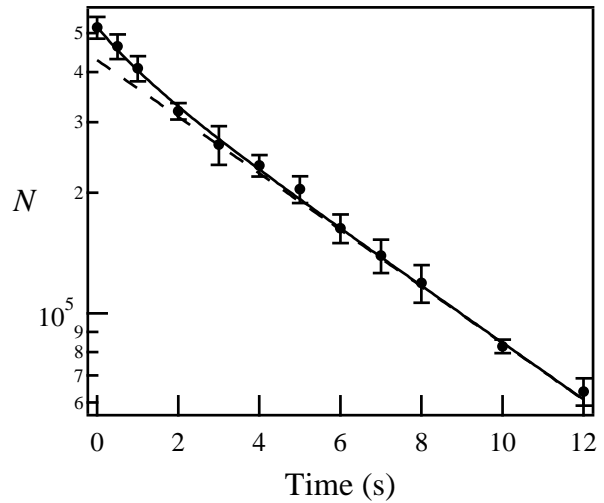}
\end{center}
\caption{Number of atoms as a function of time in the 2D lattice.
The solid line is a fit to the decay as described
in the text, and the dashed line is an extrapolation of the asymptotic 1-body loss.} \label{fig:tubeabs}
\end{figure}

Figure~\ref{fig:tubeabs} shows the total number of atoms as a
function of time $t$ in the lattice, obtained by absorption
imaging 34~ms after release from the lattice and magnetic trap. We
calibrate our absorption measurements by comparing the observed
expansion of a released condensate to
the known number-dependent expression for the expansion~\cite{castin96}. (The
inferred absorption cross section agrees ($\sim$10\%) with one
calculated from the steady state Zeeman sub-level distribution
resulting from optical pumping.) The number of atoms in the BEC
fluctuates by less than 20\% from shot to shot.  We automate the
experiment to produce a BEC every minute and each data point is
typically an average of five measurements. In order to minimize
systematic effects due to long term thermal drift of the trap
coils, we run current in the magnetic trap {\it after} the imaging
such that the total time that the magnetic trap is on is the same
for each measurement. While the atoms are in the lattice, we apply a radio frequency shield~\cite{soding99} tuned 500 kHz above the minimum of the
trap to reduce heating without significantly increasing the loss
of atoms trapped in the lattice.

We measure the size of the lattice-trapped cloud in the $xy$-plane
by phase contrast imaging. The initial column density distribution
of the cloud is well described by an integrated Thomas-Fermi (TF)
profile, with radii of $R_{x} = 13.1(5)$ $\mu$m and $R_{y} =
22.5(10)$ $\mu$m so that the observed number of atoms per tube at
$(x,y)$ is well described by ${\cal N}_{\textrm{\scriptsize tube}}
= {\cal N}_{\textrm{\scriptsize max}} (1-(x /R_x)^2-(y
/R_y)^2)^{3/2}$, where ${\cal N}_{\textrm{\scriptsize max}} = 5
N_{0} \lambda^2/8 \pi R_x R_y$ is the number of atoms in the
central tube and $\lambda/2$ is the spacing of the tubes. Based on
the initial total number and the measured sizes of the cloud, we
determine ${\cal N}_{\textrm{\scriptsize max}} = 230(40)$. In the
$xz$-plane, we measure the size of the cloud by absorption
imaging. The initial $xz$ density distribution is also described by a
TF profile, of radii $R_{x} = 15(2)$ $\mu$m (in agreement with our
phase contrast $xy$-measurement) and $R_{z} = 17(2)$ $\mu$m. For
our parameters, the atom distribution along the tubes (along z) is
not expected to deviate significantly from a TF
profile~\cite{dunjko01}; indeed $R_{z}$ agrees with the 1D TF
value calculated based on ${\cal N}_{\mbox{\scriptsize max}}$. We
note that the peak density is $\sim 1 \times 10^{15}$~cm$^{-3}$,
which would lead to rapid three-body loss in a 3D system.

We observe that the cloud slowly expands in the $z$-direction over
the course of the measurement: the cloud expands by 5~$\mu$m in 2~s while the $R_x$ and $R_y$ radii remain constant. This
expansion is consistent with a 1~kHz/s rate of energy
increase, which is less than the expected rate due to spontaneous
emission with full equilibration between the radial and
longitudinal degrees of freedom, but more than the expected rate
with no equilibration. Regardless of the mechanism, this expansion
reduces the density only modestly during the first two seconds,
when most of the three-body decay occurs. We take this into
account empirically in modelling the decay.

To model the decay using Eq.~\ref{lossrate}, we assume an overall
3D TF density profile with gaussian radial distributions
within each tube.  In addition, for simplicity of modelling we
assume $K_3^{\mbox{\scriptsize{1D}}}$ to be a constant (see below). With these
approximations, Eq.~\ref{lossrate} becomes:
\begin{equation}\label{equationdecay}
 \frac{dN}{dt}=-K_1 N-\alpha(t) K_3^{\mbox{\scriptsize{1D}}}N^{3},
\end{equation}
where $\alpha(t) = (25/896 \pi^4)
(\lambda^2/a_{\bot}^2R_{x}R_{y}R_{z}(t))^2$. The radii $R_x$ and
$R_y$ are kept constant at their measured values, and $R_z(t)$
grows linearly in time at the measured rate of 2.5~$\mu$m/s. This
differential equation has an analytic solution which gives the
total number as a function of time, to which we fit the data of Fig.~\ref{fig:tubeabs}.

With this analysis, we determine $K_3^{\mbox{\scriptsize{1D}}} = 1.2(7) \times
10^{-30}$ cm$^6$ s$^{-1}$ and $K_1 = 0.16(2)$ s$^{-1}$. This
result is relatively insensitive to the specific model used for
atomic spatial redistribution during decay, and variations among
realistic models fall within the quoted uncertainties. We
attribute $K_1$ mainly to optical pumping to states other than the
original $(1,-1)$ state, which are untrapped in the
combined optical, magnetic and gravitational potential.  For the depth and detuning of
our lattice, the majority of photon scattering events return the
atoms to the original $(1,-1)$ Zeeman sub-level and do not
contribute to $K_1$. The calculated loss rate for a 29~$E_R$
lattice is 0.17~s$^{-1}$, in good agreement with our measured
$K_1$.

\begin{figure}[t]
\begin{center}
\includegraphics[width=3.24in]{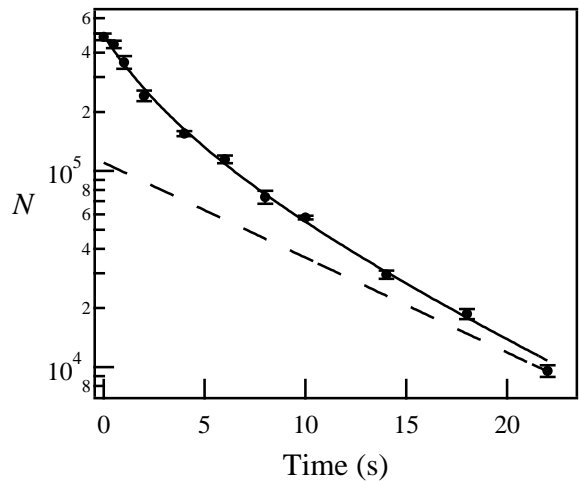}
\end{center}
\caption{Number as a function of time in the tight magnetic trap (no 2D
lattice). The solid line is a fit to the
decay~\cite{soding99}, and the dashed line is an extrapolation of the asymptotic 1-body loss.} \label{fig:becabs}
\end{figure}

To determine the reduction of three-body recombination in 1D, we
must compare $K_3^{\mbox{\scriptsize{1D}}}$ to $K_3^{\mbox{\scriptsize{3D}}}$.  A comparison in the same
apparatus reduces systematic uncertainty due to our 15\% number uncertainty. (The
effect of this systematic uncertainty is not eliminated entirely because
the power law dependence of $dN/dt$ on $N$ is different in 1D and 3D.) We therefore repeat our experiments in a
tight magnetic trap in the absence of a lattice, similar to
Refs.~\cite{burt97,soding99}. (But see~\cite{arescale}.) For the (1,-1) state, we measure~\cite{F2note} $K_3^{\mbox{\scriptsize{3D}}} = 8.3(20)\times 10^{-30}$~cm$^6$~s$^{-1}$
(see Fig.~\ref{fig:becabs}), which is in agreement with the value of $5.8(1.9) \times 10^{-30}$~cm$^6$~s$^{-1}$ measured in~\cite{burt97}.

Comparing our measurements, we find that the ratio of the three-body decay coefficients in 1D and 3D is 0.14(9). This represents a factor of seven reduction in $g^\textrm{\scriptsize 1D }_3$ over $g^\textrm{\scriptsize 3D }_3$, a clear signature of
correlations.

For comparison of the observed reduction in $g_3$ with theory, we
calculate $\gamma$ at the center of each tube.  From the experimentally determined density distribution we find at $t=0$ that $\gamma > 0.34$, with 80\%
of the atoms having $0.34< \gamma < 0.65$, and median value
$\gamma_{\mbox{\scriptsize m}}=0.45$.  We do not expect
correlations to vary significantly over this range of $\gamma$~\cite{kheruntsyan}, so the
assumption of using a single average $K_{3}^{{ \mbox{\scriptsize
1D}}}$ in the model should be reasonable.

\begin{figure}[htbp]
\begin{center}
\includegraphics[width=3.24in]{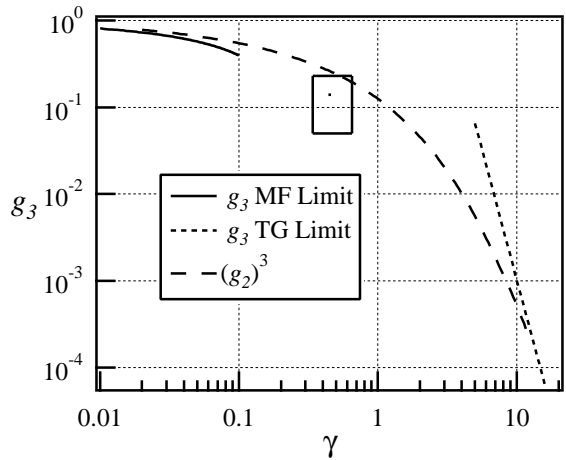}
\end{center}
\caption{Comparison of our results with theoretical calculations of the third-order correlation function
$g_3$ vs.  $\gamma$. The solid and dotted lines
represent the MF and TG limits of $g_3$ from~\cite{gangardt03},
and the dashed line is an estimate~\cite{shlyapnikov} of $g_3 =
(g_2)^3$, based on $g_2$ calculated in~\cite{kheruntsyan}. The
measured suppression factor $K_3^{\mbox{\scriptsize
1D}}/K_3^{\mbox{\scriptsize 3D}}$ is indicated by the box, where
the height represents the measurement uncertainty and the width is
the 0\% to 80\% range of $\gamma$ for our system.} \label{fig:theory}
\end{figure}

In Fig.~\ref{fig:theory} we compare the measured reduction in the
three-body loss rate coefficient with theoretical estimates for
$g_3$ in an interacting 1D gas at $T=0$. There is currently no
theory for $g_3$ in the regime intermediate to the MF and TG
limits ($\gamma \simeq 1$). We have plotted an approximate $g_3$
in the intermediate regime, using the expression $g_3 \simeq
(g_2)^3$~\cite{shlyapnikov}, and the value of $g_2$ for a
homogenous system from Ref.~\cite{kheruntsyan}. The value of $g_2$
is expected to be insensitive to $T$ for $T \ll T_d \simeq {\cal N}_{\textrm{\scriptsize max}} \hbar \omega_z /k_B $ in this range of $\gamma$.
From the measured values of ${\cal N}$ and $\omega_z$ we
estimate the distribution of degeneracy temperatures, finding at $t=0$ a peak value of $\sim$13~kHz and a median value of $\sim$9~kHz.  While it is difficult to measure
the temperature in our system, the measured size at $t=0$ is consistent with zero-temperature TF theory, and is certainly much less than $m \omega_z^2 R_z^2/2$, which during the first several seconds does not exceed $\sim 6$~kHz.

The reduction in $K_{3}^{\mbox{\scriptsize{1D}}}$ relative to $K_{3}^{\mbox{\scriptsize{3D}}}$
indicates that we are beyond the mean-field regime, and is a
signature of the ``fermionization'' of the bosonic particles in
1D.  The fundamental
effects of low dimensionality on the correlation properties of a
quantum Bose gas are also of practical interest, given the interest
in the physics of ``atom lasers'' loaded in waveguides. In
addition, these experiments indicate that high density, strongly
correlated 1D systems can be realized without fast decay due to
three-body recombination.

We acknowledge helpful conversations with M. Olshanii, G. Shlyapnikov and K. Kheruntsyan, and partial support from ARDA, ONR and NASA.

\bibliographystyle{prl}

\begin{thebibliography}{99}
\bibitem{stddev} All uncertainties reported here are one standard deviation
combined statistical and systematic uncertainties.
\bibitem{K2note} Even though correlations suppress three-body decay more than two-body decay, the upper limit for the two-body decay rate~\cite{burt97} is still negligible.
\bibitem{F2note} We also measure the three-body rate for the $(2,2)$ state,
obtaining $K_3^{3D} = 1.5(3) \times 10^{-29}$~cm$^6$/s, in
agreement with the result $1.8(4) \times 10^{-29}$~cm$^6$/s found in~\cite{soding99}. Our measurements in the (2,2) and (1,-1) states yield a ratio of 1.8(6), confirming that $K_3^{3D}$ is different for these states.
\end{thebibliography}

\end{document}